\pdfoutput=1 

\documentclass[aps,twocolumn,amsmath,amssymb,preprintnumbers,superscriptaddress,floatfix]{revtex4}
\usepackage[utf8]{inputenc}
\usepackage{newtxtext}
\usepackage[upint]{newtxmath}
\usepackage{microtype}
\usepackage{textcomp}
\usepackage{eucal}
\usepackage{bm} 

\usepackage{enumerate}
\usepackage{amsfonts}
\usepackage{amsmath}
\usepackage{amssymb}
\usepackage{color}
\usepackage{soul}

\usepackage{graphicx}

\usepackage[colorlinks,allcolors=blue]{hyperref}
\usepackage[capitalize]{cleveref}

\let\phi=\varphi
\let\epsilon=\varepsilon

\newcommand{\B}[1]{\bm{#1}}

\renewcommand{\d}[1]{\mathrm{d}{#1}}

\newcommand{\eg}{e.g.\ }
\newcommand{\ie}{i.e.\ }

\definecolor{DarkRed}{rgb}{0.80,0,0}
\definecolor{DarkBlue}{rgb}{0,0,0.80}
\definecolor{Purple}{rgb}{0.55,0,0.55}

\newcommand{\prlsection}[1]{\textit{#1}.\kern0.05em---\kern0.05em\ignorespaces}

\begin{document}
\title{Paramagnetic Meissner effect in voltage-biased proximity systems}
\author{Jabir Ali Ouassou}
\affiliation{Center for Quantum Spintronics, Department of Physics, Norwegian \\ University of Science and Technology, NO-7491 Trondheim, Norway}
\author{Wolfgang Belzig}
\affiliation{Fachbereich Physik, Universität Konstanz, D-78457 Konstanz, Germany}
\author{Jacob Linder}
\affiliation{Center for Quantum Spintronics, Department of Physics, Norwegian \\ University of Science and Technology, NO-7491 Trondheim, Norway}
\begin{abstract}
  Conventional superconductors respond to external magnetic fields by generating diamagnetic screening currents.
  However, theoretical work has shown that one can engineer systems where the screening current is \emph{paramagnetic}, causing them to \emph{attract} magnetic flux---a prediction that has recently been experimentally verified.
  In contrast to previous studies, we show that this effect can be realized in simple superconductor/normal-metal structures with no special properties, using only a simple voltage bias to drive the system out of equilibrium.
  This is of fundamental interest, since it opens up a new avenue of research, and at the same time highlights how one can realize paramagnetic Meissner effects without having odd-frequency states at the Fermi level.
  Moreover, a voltage-tunable electromagnetic response in such a simple system may be interesting for future device design.
\end{abstract}
\maketitle

%-------------------------------------------------------------------------------%
%                                 MAIN ARTICLE                                  %
%-------------------------------------------------------------------------------%

\prlsection{Introduction}
Conventional superconductors have two defining properties~\cite{Tinkham2004,Fossheim2004}.
The first is their perfect conductance of electric currents, from which they derive their name.
The second is the so-called Meissner effect, whereby dissipationless electric currents screen magnetic fields.
Both properties arise due to a coherent condensate of electron pairs (Cooper pairs) which exhibits spontaneous symmetry breaking, and it is of fundamental interest to understand both in depth.

In bulk superconductors, the Meissner effect is \emph{diamagnetic}, meaning that the screening currents try to expel magnetic flux from the superconductor.
However, the story is more complicated in \emph{proximity structures}, where superconductors and non-superconductors are combined to engineer novel device functionality.
Diamagnetic screening in such structures has been investigated some while ago \cite{Zaikin1982,Belzig:1996uh}, and several interesting impurity effects have been found~\cite{MullerAllinger:1999vw,Belzig1999}.
A cylindrical geometry can increase the diamagnetic so-called overscreening \cite{Belzig2007}.
At ultra-low temperatures, a reentrant effect was observed experimentally~\cite{VISANI:1990fi}, and even an overall paramagnetic response in thermal equilibrium \cite{MullerAllinger:2000ue}.
Other systems with unexpected properties are superconductor/ferromagnet (S/F) devices, where Cooper pairs can leak from S~to~F.
The Cooper pairs of a conventional superconductor are singlet even-frequency pairs, \ie they carry no net spin and respect time-reversal symmetry.
Once they leak into~F, some of these are converted into triplet odd-frequency pairs, which have fundamentally different properties~\cite{Eschrig2010,Blamire2014,Linder2015,Eschrig2015a,Buzdin2005,Bergeret2005,Berezinskii1974,Linder2017}.
One example is that odd-frequency pairs give rise to a \emph{paramagnetic} Meissner effect, where the screening currents \emph{attract} magnetic flux~\cite{Linder2017,Bergeret2001,Yokoyama2011,Asano2011,Mironov2012,Alidoust2014,Espedal2016}.
This effect has been predicted for a variety of S/F setups, and has been confirmed experimentally via muon-rotation experiments~\cite{DiBernardo2015}.
It has also been identified in \eg metals with repulsive electron--electron interactions~\cite{Fauchere1999} and at the interfaces of $d$-wave superconductors~\cite{Higashitani1997,Lofwander2001}.
In these systems, the effect is caused by midgap states, which are again linked to odd-frequency pairing~\cite{Linder2017}.

\begin{figure}[b!]
  \includegraphics[scale=0.85]{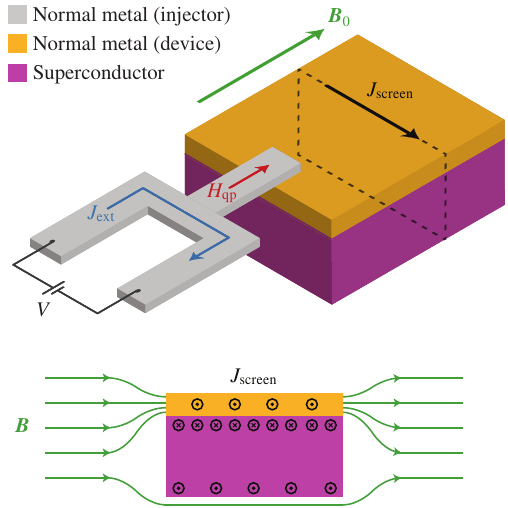}
  \caption%
  {%
    (Color online)
    Top figure: Suggested experimental setup.
    The left end features a quasiparticle injector (grey).
    The voltage source forces an electric current $J_\text{ext}$ through a normal-metal wire, causing an excess of electrons and holes to accumulate in the middle of the wire.
    This drives a diffusion $H_\text{qp}$ of excess quasiparticles onto an adjacent normal-metal film (yellow), thus driving it out of equilibrium.
    This film is also proximitized by a conventional superconductor underneath (purple), causing Andreev bound states to form there.
    The combination results in a paramagnetic effect, whereby an external magnetic field $\B{B}_0$ is \emph{enhanced} by the screening currents in the normal metal.
    Since whether the film reacts dia- or paramagnetically depends on the voltage, the device can be tuned between these Meissner responses in situ.
    Bottom figure: cross-sectional view of the device during operation, showing how the magnetic field~$\B{B}$ is deformed by the screening currents $J_\text{screen}$.
    The normal metal can have a paramagnetic response, whereby it attracts magnetic flux.
    The superconductor underneath remains diamagnetic, and therefore expels magnetic flux.
  }
  \label{fig:model}
\end{figure}

We consider a fundamentally different way of realizing the paramagnetic effect: by driving a superconductor/normal-metal (S/N) bilayer out of equilibrium via a voltage bias.
Our suggested setup is visualized in \cref{fig:model}, and explained in detail over the next two pages.
The mechanism is again related to odd-frequency superconductivity;
we will see that an essential ingredient is large subgap peaks in the normal-metal density of states (DOS), and these appear at energies where odd-frequency pairs dominate~\cite{Tanaka2007,Linder2017}.
However, our setup does not require that these reside precisely at the Fermi level (midgap states).
Instead, the Meissner response in our setup is determined by the DOS at a voltage-controlled finite energy.
Our predictions can be verified via the same setup as Ref.~\cite{DiBernardo2015}.

A related idea was discussed in Refs.~\cite{Aronov1973,Aronov1975}, where they suggested that a microwave-irradiated superconductor might become paramagnetic.
However, they concluded that the paramagnetic state would be unstable, and could therefore not be realized.
In contrast, our system avoids this instability by realizing the paramagnetic effect in a proximity system instead of a bulk system.
Moreover, voltage control may be more desirable than microwave control for potential applications.

A similar setup to ours was investigated in Ref.~\cite{Bobkova2013}, where they calculated the Meissner response of S/N structures driven out of equilibrium using a voltage-controlled quasiparticle injector.
However, they did \emph{not} find a paramagnetic response, and the main reason appears to be their parameters.
We analytically predict the effect for clean materials, thick superconductors, and low temperatures.
In contrast, Ref.~\cite{Bobkova2013} considered dirty materials, thin superconductors, and high temperatures.
This suppresses the subgap peaks in the DOS, which we will see are essential for a paramagnetic response.

\prlsection{Motivation}
Consider a superconducting system that is exposed to a weak magnetic field~$\B{B} = \nabla\times\B{A}$, which we describe via a vector potential~$\B{A}$.
For concreteness, let us consider a geometry where a thin film at $0<z<d$ is subjected to a magnetic field~$\B{B} \sim \B{e}_y$, which we describe via the vector potential $\B{A} = A(z)\,\B{e}_x$.
This is identical to the experimental geometry employed in Ref.~\cite{DiBernardo2015}.
In the clean and nonlocal limits, the linear-response screening current is then given by
\begin{equation}
  \B{J}(z) = -K\langle \B{A} \rangle_z = -\frac{K}{d} \int_0^d \d{z'}\,\B{A}(z').
  \label{eq:screening-current}
\end{equation}
Here, we have introduced the screening kernel
\begin{equation}
  K = \frac{1}{3} e^2 v_\text{F}^2 \left[ N_\text{F} - \int_{-\infty}^{+\infty} \d{\epsilon} \, N(\epsilon) \left(-\frac{\partial f}{\partial \epsilon}\right) \right],
  \label{eq:screening-kernel}
\end{equation}
where $\epsilon$ is the quasiparticle energy, $f(\epsilon)$ the distribution function, $N(\epsilon)$ the DOS, $N_\text{F}$ the Fermi-level DOS in the non-superconducting state, $v_\text{F}$ the Fermi velocity, and $e$ the electron charge.
These equations are derived in standard textbooks on superconductivity~\cite{Tinkham2004,Schrieffer1999}, and have previously been used to \eg predict paramagnetic effects in materials with repulsive electron interactions~\cite{Fauchere1999}, $d$-wave superconductors~\cite{Higashitani1997}, and microwave-irradiated superconductors~\cite{Aronov1973,Aronov1975}.
We provide a simple and compact derivation of this equation within the quasiclassical formalism in the Supplemental Material.

Many well-known results for Meissner effects can be seen directly from \cref{eq:screening-current,eq:screening-kernel}.
In equilibrium, the distribution has a Fermi--Dirac form, which at low temperatures reduces to a step function ${f(\epsilon) \approx \theta(-\epsilon)}$.
Substituted into \cref{eq:screening-kernel}, this produces the simplified equation ${K \sim N_\text{F} - N(0)}$.
For a BCS superconductor, there is a gap around the Fermi level~${\epsilon = 0}$, and ${N(0) = 0}$ causes ${K > 0}$.
This produces a diamagnetic response.
On the other hand, in systems with odd-frequency pairing, one can have a zero-energy peak in the DOS, and ${N(0) > N_\text{F}}$ causes~${K < 0}$.
This produces a paramagnetic response.

We are interested in a new way to realize the paramagnetic Meissner effect: by manipulating the distribution~$f(\epsilon)$ instead of the DOS~$N(\epsilon)$.
Before we discuss its exact physical origin, let us just assume that one can induce a two-step Fermi--Dirac-like distribution, which at low temperatures reduces to
\begin{equation}
  \label{eq:distribution-general}
  f(\epsilon) \approx [\theta(+\Omega-\epsilon) + \theta(-\Omega-\epsilon)]/2.
\end{equation}
We note that the effect of $\Omega$ is essentially to excite electrons in the range $0 < \epsilon < \Omega$ and holes in the range $-\Omega < \epsilon < 0$, resulting in an \emph{excited energy mode} or \emph{increased effective temperature}.
Substituting the above into \cref{eq:screening-kernel}, and using the electron--hole symmetry of the DOS $N(+\epsilon) = N(-\epsilon)$, we get
\begin{equation}
  K = \frac{1}{3}  e^2 v_\text{F}^2 \left[ N_\text{F} - N(\Omega) \right].
  \label{eq:screening-kernel-omega}
\end{equation}
In other words, if we can tune $\Omega$, it is now sufficient that $N(\epsilon) > N_\text{F}$ at \emph{some} energy~$\epsilon$ for us to realize a paramagnetic state.
For example, consider the DOS of a BCS superconductor,
\begin{equation}
  N(\epsilon) = N_\text{F} \frac{|\epsilon|}{\sqrt{\epsilon^2 - \Delta^2}} \theta(|\epsilon|-\Delta).
  \label{eq:dos-superconductor}
\end{equation}
Clearly, the step function indicates that $N(\Omega) = 0$ within the gap $|\Omega| < \Delta$, resulting in a purely diamagnetic response there.
However, if we can increase its value to $|\Omega| > \Delta$, suddenly we find that $N(\Omega) \gg N_\text{F}$ due to the BCS coherence peaks, resulting in a strong paramagnetic response instead.
It would therefore be interesting if we could find a system where $\Omega$ could be tuned in situ, making it possible to actively toggle between diamagnetic and paramagnetic Meissner effects in a device.

\prlsection{Model system}
One way to realize the distribution in \cref{eq:distribution-general} is to voltage bias a normal-metal wire.
At low temperatures, the distributions at the two ends of the voltage source are just $f_\pm(\epsilon) \approx \theta(\pm eV/2 - \epsilon)$, which we use as our boundary conditions.
If the wire is short compared to the inelastic scattering length of the material, which diverges at low temperatures~\cite{Black2000}, the Boltzmann equation for the distribution reduces to a Laplace equation $\nabla^2 f = 0$~\cite{Pothier1997}.
Near the center of the wire, the solution to these equations is just $f = (f_+ + f_-)/2$~\cite{Pothier1997}.
In other words, this allows us to realize \cref{eq:distribution-general}, where $\Omega = eV/2$ is a voltage-tunable control parameter.
This result is robust to the presence of superconductivity and for resistive interfaces~\cite{Voltage,Keizer2006}.

If the center of such a wire is now connected to a different material, the wire functions as a \emph{quasiparticle injector}.
Essentially, the electrons and holes that are excited in the normal-metal wire diffuse into the adjacent material, thus inducing the distribution $f=(f_+ + f_-)/2$ there as well.
We note that this is just one way to excite a distribution like in \cref{eq:distribution-general}. 
Other alternatives that may be experimentally relevant include applying the voltage bias directly to the other material via tunneling contacts~\cite{Voltage}, or using microwaves to excite the quasiparticles~\cite{Aronov1973,Aronov1975}.
We should also note that \cref{eq:distribution-general} has previously been shown to induce many other interesting effects in superconducting systems \cite{Volkov1995,Wilhelm1998,Baselmans1999,Belzig1999,Wilhelm2000,Heikkila2000,Yip2000,Chtchelkatchev2002,Keizer2006,Bobkova2016,Voltage,Exchange,Snyman2009,Moor2009}, including a superconducting transistor \cite{Volkov1995,Wilhelm1998,Baselmans1999,Belzig1999,Wilhelm2000}, and a loophole in the Chandrasekhar--Clogston limit~\cite{Voltage}.

If we could simply connect the quasiparticle injector to a BCS superconductor, the combination of \cref{eq:dos-superconductor,eq:screening-kernel-omega} should have a paramagnetic response for voltages $eV/2 > \Delta$.
Unfortunately, for such large voltages, the superconducting state becomes energetically unfavourable \cite{Keizer2006,Snyman2009,Moor2009,Voltage}, a phenomenon that is intimately related to Chandrasekhar--Clogston physics \cite{Moor2009,Voltage,Chandrasekhar1962,Clogston1962}.
The solution is to consider S/N proximity systems, where we can produce peaks with $N(\epsilon) > N_\text{F}$ at subgap energies $\epsilon < \Delta/2$.
Note that these peaks correspond to energies where odd-frequency pairing dominates~\cite{Tanaka2007,Linder2017}.
In this way, we can induce a paramagnetic response in~N, while S remains diamagnetic and stable.
\Cref{fig:model} visualizes the experimental setup suggested based on the arguments above.
%We note that the local Meissner response can be probed via muon-spin spectroscopy experiments, using the same experimental setup as in Ref.~\cite{DiBernardo2015}.

For concreteness, let us take S to lie in $-\infty < z < 0$, and N to lie in $0 < z < d$.
The system is assumed to be infinite and translation-invariant in the $xy$-plane.
Furthermore, to make analytical progress, let us assume that there is a negligible inverse proximity effect so that $\Delta(z) \approx \Delta_0 \theta(-z)$, that the S/N interface at $z=0$ is completely transparent, that the normal-metal/vacuum interface at $z=d$ is specularly reflecting, and that the materials are clean.
In these limits, the DOS in S is just given by \cref{eq:dos-superconductor}.
In~N, however, the DOS has Andreev bound states below the gap, which for $\epsilon \ll \Delta_0$ produces the DOS:
\begin{equation}
  N(\epsilon) = N_\text{F} (\epsilon/2\epsilon_\text{A})\, \psi_1(\lfloor \epsilon/2\epsilon_\text{A} + 1/2 \rfloor + 1/2),
  \label{eq:normalmetal-dos}
\end{equation}
where the Andreev energy $\epsilon_\text{A} = \pi v_\text{F}/4d$ and $\psi_1$ is the trigamma function.
We provide a complete derivation of this result within the quasiclassical formalism in the Supplemental Material.
This result was originally derived via the Bogoliubov--de\,Gennes formalism in Ref.~\cite{deGennes1963};
their results are identical to ours in the limit $\epsilon \ll \Delta_0$ if we use the series representation of the polygamma function.
It is worth noting that for $\epsilon < \epsilon_\text{A}$, the result is just linear: $N(\epsilon) = N_\text{F} \,(\pi^2/4)\,(\epsilon/\epsilon_\text{A})$.

\begin{figure}[tb!]
  \includegraphics[width=0.88\columnwidth]{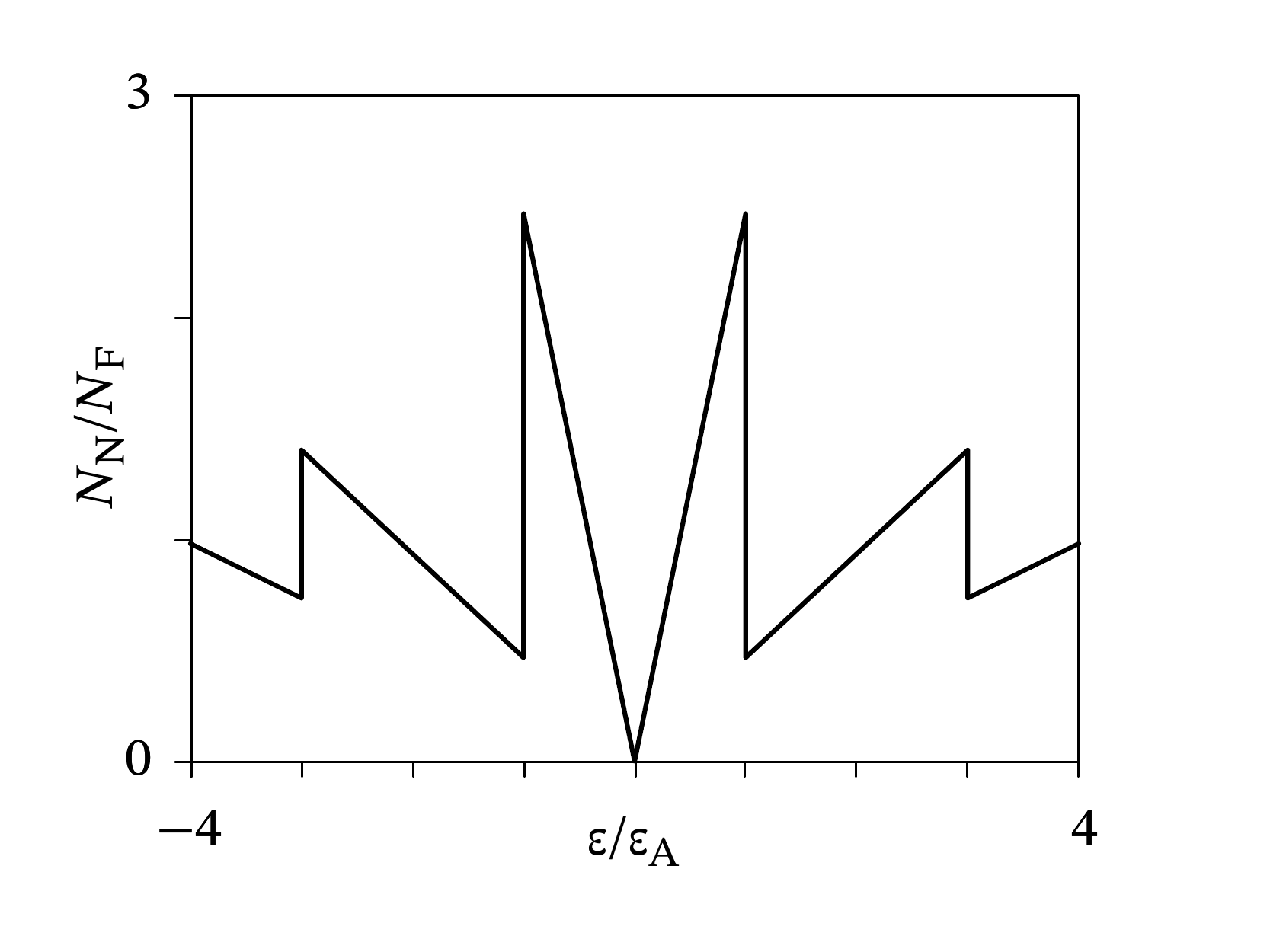}\\[-3ex]
  \caption%
  {%
    Density of states in the normal metal.
    The energy is normalized to the Andreev energy, which for \eg $d=3\xi$ would be $\epsilon_\text{A} \approx \Delta/4$.
    Note that the peaks where $N > N_\text{F}$ correspond to energies where odd-frequency Cooper pairs dominate in the normal metal~\cite{Tanaka2007,Linder2017}.
  }
  \label{fig:DOS}
\end{figure}

Let us now consider the screening kernels in this proximity system using \cref{eq:screening-kernel-omega} with $\Omega = eV/2$ and the densities of states derived above.
In~S, we have already established that $N(\Omega) = 0$ yields a purely diamagnetic response.
This is usually described via the magnetic penetration depth $\lambda_0 = 1/\sqrt{K}$,
\begin{equation}
  K = \frac{1}{\lambda_0^2} = \frac{1}{3} e^2 v_\text{F}^2 N_\text{F}.
\end{equation}
\Cref{eq:normalmetal-dos} gives a more interesting expression,
\begin{equation}
  K = \frac{1}{\lambda_0^2} [1 - (V/2V_\text{A})\,\psi_1(\lfloor V/2V_\text{A} + 1/2 \rfloor + 1/2)],
  \label{eq:screening-kernel-normal}
\end{equation}
where we reused the penetration depth~$\lambda_0$ defined in~S, and introduced the Andreev voltage~$V_\text{A} = 2\epsilon_\text{A}/e = \pi v_\text{F}/2ed$.

\begin{figure}[tb!]
  \includegraphics[width=0.88\columnwidth]{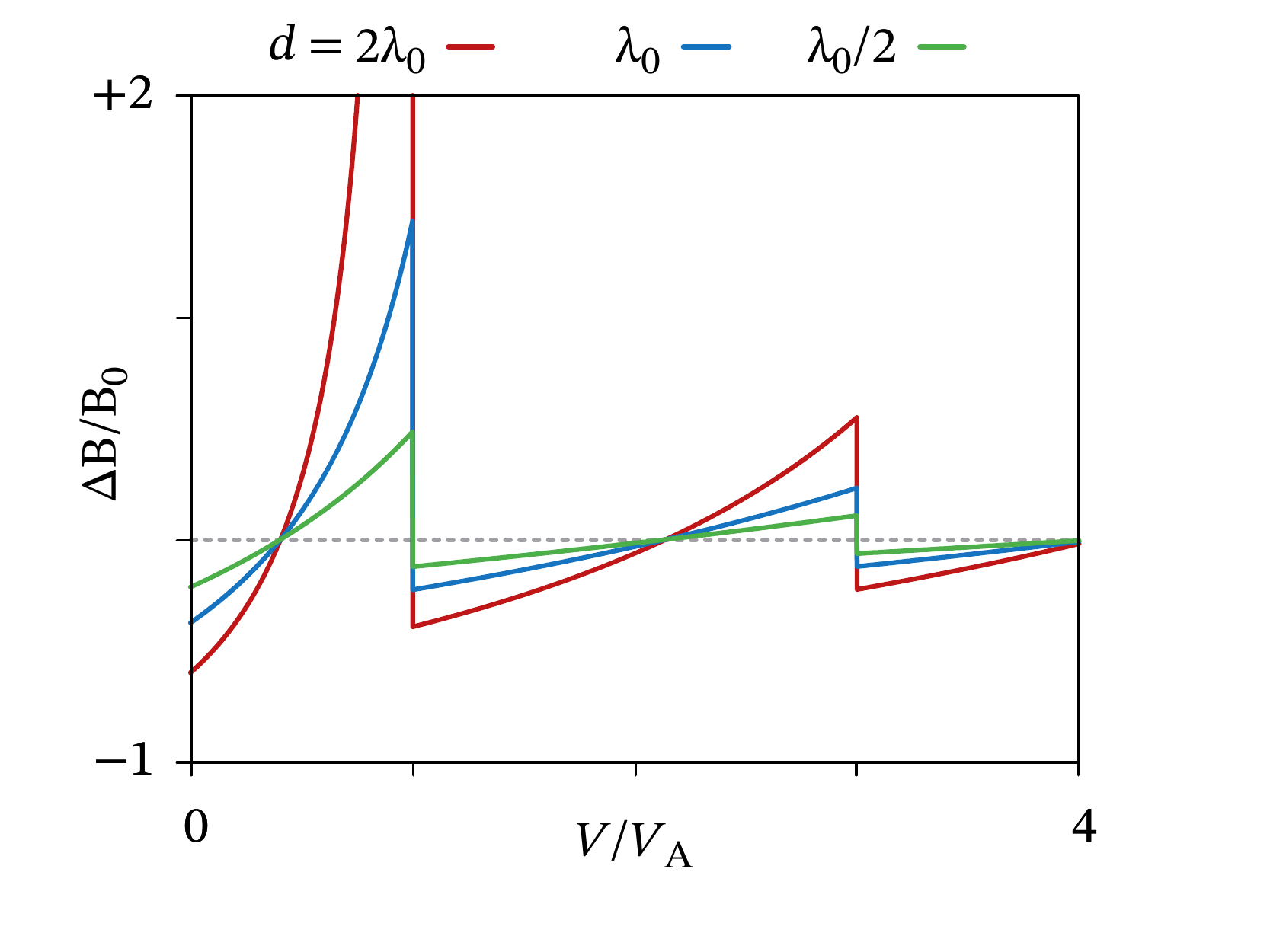}\\[-3ex]
  \caption%
  {%
    Magnetic shift in the normal metal as function of the applied voltage.
    Different curves correspond to different thicknesses~$d$ of the normal metal.
    Note that when $d > \lambda_0$, the paramagnetic effect produces a field~$B_0 + \Delta B$ many times larger than the applied field~$B_0$.
  }
  \label{fig:shift}
\end{figure}

Now that we have an expression for the screening kernel~$K$, we can solve the Maxwell equation~$-\nabla^2\B{A} = \B{J}$ together with the screening equation $\B{J} = -K\langle \B{A} \rangle_z$.
As boundary conditions, we have $\nabla\times\B{A}(d) = \B{B}_0$ at the vacuum boundary, and $\B{A}(-\infty) = 0$ deep inside~S.
We have considered a geometry where we can write $\B{A} = A(z)\,\B{e}_x$, which means that the applied magnetic field $\B{B}_0 = A'(d)\, \B{e}_y$.
As an approximation, one might also set $\B{A}(0) \approx 0$ to make analytical progress, meaning that S is assumed to perfectly screen fields near its interface.
Thus, the equations for the gauge field can be written
\begin{equation}
\begin{aligned}
  A''(z) &= K\langle A \rangle_z, \\
  A'(d)  &= B_0, \\
  A(0)   &= 0.
\end{aligned}
\end{equation}
The solution to the differential equation is $A(z) = az^2 + bz + c$ with $a = K \langle A \rangle_z/2$.
The boundary conditions then provide the constraints $b = B_0 - 2ad$ and $c = 0$.
Together, these yield
\begin{equation}
  A(z) = K\langle A \rangle_z (z^2/2 - zd) + B_0z.
  \label{eq:gauge-result}
\end{equation}
We can then calculate the average $\langle A \rangle_z$.
Using the moments $\langle z \rangle_z = d/2$ and $\langle z^2 \rangle_z = d^2/3$, and solving for $\langle A \rangle_z$, we find
\begin{equation}
  \langle A \rangle_z = \frac{B_0d/2}{1 + Kd^2/3}.
  \label{eq:gauge-avg}
\end{equation}
We now go back to \cref{eq:gauge-result} to calculate the magnetic field,
\begin{equation}
  B(z) = A'(z) = B_0 + K\langle A \rangle_z(z-d).
\end{equation}
Substituting in \cref{eq:gauge-avg} into this result, we obtain an analytical result for the magnetic field inside~N:
\begin{equation}
  B(z) = B_0 \left( 1 + \frac{K(z-d)(d/2)}{1 + Kd^2/3} \right).
\end{equation}
The net magnetic field change~$\Delta B = B(0) - B(d)$ induced by the screening currents can then be calculated as
\begin{equation}
  \frac{\Delta B}{B_0} = -\frac{Kd^2/2}{1 + Kd^2/3}.
\end{equation}
To obtain the final results, we just have to substitute in \cref{eq:screening-kernel-normal}:
\begin{equation}
\begin{aligned}
  \frac{\Delta B}{B_0}     \;=\;&        -\frac{\rho/2}{\lambda_0^2/d^2 + \rho/3}, \\
  \rho(V)                  \,=\;&   1 - (V/2V_\text{A}) \,\psi_1(\lfloor V/2V_\text{A}+1/2\rfloor + 1/2).
  \label{eq:mainresult}
\end{aligned}
\end{equation}
This provides us with a simple analytical result for the linear response $\Delta B$ of a clean proximitized metal to an applied field~$B_0$.
The result is expressed in terms of the Andreev voltage $V_\text{A} = \pi v_\text{F}/2ed$.
This can be put into more familiar terms by introducing the superconducting coherence length $\xi = v_\text{F}/\Delta_0$; for instance, an N of length $d = 3\xi$ would yield $V_\text{A} \approx \Delta_0/2e$.
This magnetic shift as function of voltage is shown in \cref{fig:shift}.

\prlsection{Discussion}
In the previous sections, we have shown that a paramagnetic Meissner effect can arise in the device in \cref{fig:model}.
Moreover, we have shown that whether the response is diamagnetic or paramagnetic can be controlled using a voltage.
Our main result is \cref{eq:mainresult}, which provides a simple analytical solution for the magnetic field shift~$\Delta B$ that occurs for a given external magnetic field~$B_0$ and voltage~$V$.
These predictions can be tested using a muon-rotation experiment to directly probe the local magnetic field at different points inside the device, using basically the same setup as in Ref.~\cite{DiBernardo2015}.

The striking results are shown in \cref{fig:shift}, where we see that for sufficiently thick normal metals, $\Delta B$ appears to diverge as the voltage~$V$ approaches the Andreev voltage~$V_\text{A}$.
Since we considered a completely clean material at zero temperature, there is an abrupt transition between paramagnetism and diamagnetism as the voltage is increased beyond the Andreev voltage.
In realistic systems, such sharp features should be smeared out by finite temperature, elastic scattering, and inelastic scattering.

We have also assumed perfect transparency at the S/N interface. 
Since a finite interface resistance can be expected to dampen the resonance peaks in the DOS of N, we would expect the paramagnetic Meissner effect to become weaker for opaque interfaces.
We also note that in regions where $\Delta B \gg B_0$, a linear-response calculation is not technically valid anymore, and a full nonlinear-response calculation is warranted if one requires quantitatively rigorous results.
Nevertheless, we would expect our results to remain qualitatively valid in such systems, and investigating this rigorously would be an interesting avenue for further research.
For instance, whether a paramagnetic instability that generates a spontaneous magnetic flux can appear requires a nonlinear-response calculation~\cite{Fauchere1999}.

Another interesting proposition for further research, would be to investigate whether a paramagnetic Meissner effect can be induced in dirty systems as well.
While no such effect was detected in Ref.~\cite{Bobkova2013}, they focused on high temperatures and thin superconductors, while the opposite limit may be the relevant one.
Even if it is not possible to realize the effect in a dirty S/N structure, the prospect of a voltage-controllable Meissner effect in dirty S/F devices may still be of interest.

\prlsection{Conclusion}
Using a simple linear-response calculation, we have demonstrated how nonequilibrium effects can give rise to a paramagnetic Meissner response.
Moreover, we have provided a specific experimental proposal where the magnetic response can be controlled in situ via an applied voltage.
In addition to being relevant to the fundamental study of the Meissner effect and odd-frequency superconductivity, our results demonstrate a new way to control the interaction between superconducting structures and magnetic fields via nonequilibrium effects.
This paves the way for a new line of research in this direction, and may be relevant to future superconducting device design.

%-------------------------------------------------------------------------------%
%                                ACKNOWLEDGEMENTS                               %
%-------------------------------------------------------------------------------%

\vspace{3ex}
\begin{acknowledgments}
  J.A.O. and J.L. were supported by the Research Council of Norway through grant 240806, and its Centres of Excellence funding scheme grant 262633 ``\emph{QuSpin}''. W.B. thanks Arne Brataas and the Centre of Excellence ``\emph{QuSpin}'' for hospitality and the DFG through SFB 767 for financial support.
\end{acknowledgments}

%-------------------------------------------------------------------------------%
%                                  BIBLIOGRAPHY                                 %
%-------------------------------------------------------------------------------%

% Bibtex file
\bibliography{references}

% Bibtex refs
% [include bbl-file here]

\end{document}

\title{Supplemental Material}
\author{Jabir Ali Ouassou}
\affiliation{Center for Quantum Spintronics, Department of Physics, Norwegian \\ University of Science and Technology, NO-7491 Trondheim, Norway}
\author{Wolfgang Belzig}
\affiliation{Fachbereich Physik, Universität Konstanz, D-78457 Konstanz, Germany}
\author{Jacob Linder}
\affiliation{Center for Quantum Spintronics, Department of Physics, Norwegian \\ University of Science and Technology, NO-7491 Trondheim, Norway}
\begin{abstract}
  In \cref{sec:dos}, we consider a clean superconductor/normal-metal bilayer with a transparent interface, and calculate the subgap density of states.
  This reproduces the well-known solution by de\,Gennes and Saint-James~\cite{deGennes1963}.
  In \cref{sec:linear-response}, we calculate the linear response of a superconducting material to a magnetic field.
  This reproduces the standard textbook result~\cite{Tinkham2004,Schrieffer1999}.
  Both our derivations are based on the quasiclassical Eilenberger formalism.
\end{abstract}
\maketitle

\section{Density of states}\label{sec:dos}
In the quasiclassical and clean limits, superconductivity in S/N systems can be described by the Eilenberger equation~\cite{Eilenberger1968,Belzig1999}
\begin{equation}
  \label{eq:eilenberger}
  -i\B{v}\cdot\tilde{\nabla} \H{g}^R = [\epsilon \H{\tau}_3 + \Delta\,i\H{\tau}_1, \hat{g}^R],
\end{equation}
where $\tilde{\nabla} \H{g}^R = \nabla \H{g}^R - ie\B{A} [\H{\tau}_3, \H{g}^R]$ is a gauge-covariant derivative, $\H{g}^R$ is the retarded propagator, $\epsilon$ is the quasiparticle energy, and $\B{v}$ is a velocity on the Fermi surface.
All quantities with hats have a $2\times2$ structure in Nambu space, and $\H{\tau}_n$ are the standard Pauli matrices.
We take the order parameter~$\Delta$ to be a real constant in superconductors and zero in normal metals.

Let us first consider the case of a bulk superconductor in the absence of magnetic fields.
The solution can then be written
\begin{equation}
  \label{eq:bulk-prop}
  \H{g}^R_0 = \frac{1}{\Omega} [\epsilon \H{\tau}_3 + \Delta \,i\H{\tau}_1].
\end{equation}
It is straight-forward to verify that this satisfies \cref{eq:eilenberger} in the absence of inhomogeneities and magnetic fields.
Up to a sign~$\eta$, the prefactor follows from the normalization ${(\H{g}^R)^2 = 1}$,
\begin{equation}
  \Omega = \eta \, \sqrt{\epsilon^2 - \Delta^2}.
\end{equation}
Finally, the sign is then fixed by the fundamental symmetries $g_{11}(+\epsilon,+\B{v}) = -g_{22}^*(-\epsilon,-\B{v})$ and $g_{12}(+\epsilon,+\B{v}) = g_{21}^*(-\epsilon,-\B{v})$,
\begin{equation}
  \eta = 
  \begin{cases}
    1 & \text{for $|\epsilon| < \Delta$,} \\
    \sgn\epsilon & \text{for $|\epsilon| > \Delta$}.
  \end{cases}
\end{equation}
Note that $\Omega$ is imaginary for $|\epsilon| < \Delta$ so that $\Omega^*(-\epsilon) = -\Omega(+\epsilon)$.
From the above, we can easily reproduce the density of states of a bulk superconductor $N_0(\epsilon) = (N_F/2) \re\tr[\H{\tau}_3\H{g}^R_0]$:
\begin{equation}
  \label{eq:bulk-dos}
  N_0(\epsilon) = N_F \frac{|\epsilon|}{\sqrt{\epsilon^2 - \Delta^2}} \Theta(|\epsilon| - \Delta).
\end{equation}

We now consider a clean S/N bilayer, which consists of a thick superconductor at $-\infty < z < 0$ coupled to a thin normal metal at $0 < z < d$.
Deep inside the superconductor $z \rightarrow -\infty$, the propagators are expected to converge to the bulk solution~$\H{g}^R_0$ from \cref{eq:bulk-prop}.
At the vacuum interface $z = d$, we assume specular reflection $\H{g}^R(\epsilon, +v_z, d) = \H{g}^R(\epsilon, -v_z, d)$.
Finally, we take the interface at $z = 0$ to be transparent, and approximate the order parameter by a step function~$\Delta(z) = \Delta \theta(-z)$.

We will focus on subgap energies below, meaning that $\Omega = i\kappa = i\sqrt{\Delta^2 - \epsilon^2}$ is imaginary.
The bulk solution is then
\begin{equation}
  \H{g}^R_0 = \kappa^{-1} [-i\epsilon \H{\tau}_3 + \Delta \H{\tau}_1].
\end{equation}
As for the full solutions $\H{g}^R_S$ and $\H{g}^R_N$ in the superconductor and normal metal, respectively, these can be written:
\begin{align}
  \H{g}^R_S &= \H{g}^R_0 + A \kappa^{-1} e^{2\kappa z/|v_z|} [-i\Delta\H{\tau}_3 + \epsilon \H{\tau}_1 + \kappa\H{\tau}_2\sgn v_z], \label{eq:prop-S} \\
  \H{g}^R_N &= B\H{\tau}_3 + C \H{\tau}_1 \cos[2\epsilon(z-d)/|v_z|] \label{eq:prop-N} \\
            &\;\phantom{= B \H{\tau}_3} - C\H{\tau}_2 \hspace{0.075em} \sin[2\epsilon(z-d)/|v_z|] \sgn v_z. \notag
\end{align}
These solutions have a similar form to those found using the Matsubara formalism in \eg Refs.~\cite{Zaikin1982,Belzig2007,Reeg2014}.
The solutions themselves are easily verified by substitution into the field-free Eilenberger equation $-iv_z \partial_z \H{g}^R = [\epsilon \H{\tau}_3 + \Delta\theta(-z), \H{g}^R]$.
The solution in the superconductor has been chosen such that $\H{g}^R_S \rightarrow \H{g}^R_0$ as $z \rightarrow -\infty$, and a straight-forward calculation shows that it also satisfies the normalization condition.
In the normal metal, the specular reflection at the vacuum interface requires the odd-frequency odd-parity component to vanish, which is ensured by the sine-function.
Thus, the above is a general solution of the Eilenberger equation for a clean S/N bilayer, assuming that the superconductor is infinitely thick while the normal metal has a specularly reflecting edge.
%We have not yet utilized the boundary conditions at~$z = 0$.

Comparing corresponding Pauli matrices at $z = 0$, the transparent boundary conditions yield the constraints
\begin{align}
  A &= C \sin(2\epsilon d/|v_z|), \\
  \Delta + A\epsilon &= \kappa C \cos(2\epsilon d/|v_z|), \\
  \epsilon + A\Delta &= i\kappa B.
\end{align}
From the first two coefficient equations above, we find that
\begin{equation}
  A = \frac{\Delta}{\kappa \cot(2\epsilon d/|v_z|) - \epsilon}.
\end{equation}
Combined with the last coefficient equation, we find
\begin{equation}
  B = i\left( \frac{\epsilon  \cot(2\epsilon d/|v_z|) + \kappa}{\epsilon - \kappa \cot(2\epsilon d/|v_z|)} \right).
  \label{eq:Bcoeff}
\end{equation}
This is consistent with the analytical continuation $\omega \rightarrow -i\epsilon$ and $\Omega \rightarrow \kappa$ of Eq.~(10) in Ref.~\cite{Zaikin1982} or Eq.~(16) in Ref.~\cite{Belzig2007}.

Before calculating any physical observables, an important detail is that we have to let $\epsilon \rightarrow \epsilon+i\delta$, where $\delta\rightarrow0^+$ is a positive infinitesimal.
Since $\sgn\delta$ is the difference between the retarded and advanced propagators, this is required to obtain the correct solution branch.
For brevity, we also introduce the auxiliary quantities $\epsilon' = 2\epsilon d/|v_z|$ and $\delta' = 2\delta d/|v_z|$.
Finally, using the addition rule for the cot function, we find that ${\cot(\epsilon'+i\delta')} \approx {\cot\epsilon'} - {i\delta'\csc^2\epsilon'}$.
Expanding both the numerator and denominator to leading order in $\delta$, we find:
\begin{equation}
  B = i\left( \frac{(\epsilon\cot\epsilon'+\kappa)+i(\delta\cot\epsilon'-\delta'\epsilon\csc^2\epsilon')}{(\epsilon-\kappa\cot\epsilon')+i(\delta+\delta'\kappa\csc^2\epsilon')} \right).
\end{equation}
Clearly, the infinitesimal terms in the numerator will simply vanish in the limit $\delta\rightarrow0$, and may therefore be discarded.
The remaining infinitesimals in the denominator may be redefined as another infinitesimal $\delta'' = \delta + \delta'\kappa\csc^2\epsilon$.
This yields
\begin{equation}
  B = \frac{i(\epsilon\cot\epsilon'+\kappa)}{(\epsilon-\kappa\cot\epsilon')+i\delta''}.
\end{equation}
We then expand the fraction by the complex conjugate of the denominator, and take the real part of the result, and find
\begin{equation}
  \re B = \frac{(\epsilon\cot\epsilon'+\kappa)\delta''}{(\epsilon-\kappa\cot\epsilon')^2+(\delta'')^2}.
\end{equation}
We may then take the limit $\delta''\rightarrow0$, and write the result in terms of a Dirac delta function ${\delta''/[f(\epsilon)^2+(\delta'')^2]} \rightarrow {\pi\delta[f(\epsilon)]}$.
Reinstating $\epsilon' = 2\epsilon d/|v_z|$, the result can be written
\begin{equation}
  \re B = \pi\,[\epsilon\cot(2\epsilon d/|v_z|)+\kappa]\, \delta[\epsilon-\kappa\cot(2\epsilon d/|v_z|)].
\end{equation}
This is an exact solution at all subgap energies~$\epsilon<\Delta$.
To make analytical progress, we now focus on the low-energy limit $\epsilon \ll \kappa \approx \Delta$, in which case the above simplifies to
\begin{equation}
  \re B = \pi\delta[\cot(2\epsilon d/|v_z|)].
  \label{eq:ReB}
\end{equation}
The zeros of $\cot(2\epsilon d/|v_z|)$ are the discrete energies
\begin{equation}
  \begin{aligned}
    \epsilon_n &= \frac{\pi |v_z|}{2d} (n+1/2), & n&\in\mathbb{Z}.
  \end{aligned}
\end{equation}
The delta function in \cref{eq:ReB} can then be expanded using the identity $\delta[f(\epsilon)] = \sum_n |f'(\epsilon_n)|^{-1} \delta(\epsilon-\epsilon_n)$, 
\begin{equation}
  \re B = \frac{\pi|v_z|}{2d} \sum_n\delta(\epsilon-\epsilon_n).
\end{equation}

We now calculate the low-energy density of states~$N_N(\epsilon) = (N_F/2) \re\tr\langle\H{\tau}_3 \H{g}^R_N\rangle = N_F \langle \re B \rangle$ in the normal metal, where the angle brackets denote an average over the Fermi surface.
First, it is convenient to parametrize the result above in terms of the transport direction $v_z = v_F \cos \theta$.
Exploiting the symmetries of the result, the angular average can be written
\begin{equation}
  \frac{N_N(\epsilon)}{N_F}  = 2\epsilon_A \sum_n \int_0^1 \d{(\cos\theta)} \cos\theta\, \delta[\epsilon-\epsilon_n(\theta)],
\end{equation}
where we have introduced the Andreev energy~$\epsilon_A = \pi v_F/4d$.
The contents of the delta function can be written out as
\begin{equation}
  \epsilon - \epsilon_n(\theta) = 2\epsilon_A(n+1/2) \left( \frac{\epsilon}{2\epsilon_A(n+1/2)} - \cos\theta \right).
\end{equation}
Since we will only integrate over $0 < \cos\theta < 1$, clearly this can only become zero for $n > \epsilon/2\epsilon_A - 1/2 > 0$, which for integral values of $n$ implies that $n \geq n_A = \lfloor \epsilon/2\epsilon_A + 1/2 \rfloor$.
Using $\delta[f(\cos \theta)] = \sum_n |f'(\cos\theta_n)|^{-1} \delta(\cos\theta-\cos\theta_n)$, and then performing the integral over~$\cos\theta$, we find the result
\begin{equation}
  \frac{N_N(\epsilon)}{N_F}  = \frac{\epsilon}{2\epsilon_A} \sum_{n = n_A}^{\infty} \frac{1}{(n+1/2)^2}.
\end{equation}
This reproduces the low-energy limit of the classic Ref.~\cite{deGennes1963}.
Compared to the series expansion of polygamma functions,
\begin{equation}
  \psi_n(w) = \sum_{k=0}^\infty \frac{(-1)^{n+1} n!}{(w+k)^{n+1}},
\end{equation}
we see that the result is simply proportional to a trigamma function~$\psi_1(n_A+1/2)$.
Finally, substituting back the definition of~$n_A$, we find an explicit analytical result for the density of states in the normal metal at low energies:
\begin{equation}
  N_N(\epsilon) = N_F(\epsilon/2\epsilon_A)\,\psi_1(\lfloor \epsilon/2\epsilon_A + 1/2 \rfloor + 1/2).
  \label{eq:DOS-final}
\end{equation}
Note that below the Andreev energy~$\epsilon_A = \pi v_F/4d$, the trigamma factor is $\psi_1(1/2) = \pi^2/2$, resulting in a linear density of states $N_N = N_F(\pi^2/4)(\epsilon/\epsilon_A) \approx 2.5N_F(\epsilon/\epsilon_A)$.
At the Andreev energy, it abruptly drops from $\sim\!2.5N_F$ to $\sim\!0.5N_F$.

The analytical result above was derived under the assumption that $\epsilon \ll \Delta$.
Since we are interested in obtaining $N(\epsilon) > N_F$ in order to observe a paramagnetic effect, and we found that this in practice requires energies up to roughly the Andreev energy~$\epsilon_A$, we should in practice consider systems where $\epsilon_A \ll \Delta$.
In terms of the coherence length~$\xi = v_F/\Delta$, the Andreev energy can be written $\epsilon_A = (\pi/4)(\xi/d)\Delta$.
Thus, we require a somewhat long normal metal, and choosing \eg $d = 3\xi$ gives $\epsilon_A \approx \Delta/4$.

\section{Linear response}\label{sec:linear-response}
By reorganizing the terms in \cref{eq:eilenberger}, we see that the effect of applying a field~$\B{A}$ is actually similar to an energy shift,
\begin{equation}
  -i\B{v}\cdot\nabla \H{g}^R = [(\epsilon + e\B{v}\cdot\B{A}) \H{\tau}_3 + \Delta\,i\H{\tau}_1, \hat{g}^R],
\end{equation}
This implies that if we already know the solution~$\H{g}_0^R(\epsilon, \B{v}, \B{r})$ without a field~$\B{A}$, then the effect of a applying a field is to shift the solution to $\hat{g}^R = \hat{g}_0^R(\epsilon + e\B{v}\cdot\B{A}, \B{v}, \B{r})$.
We have assumed the long-wavelength limit here, meaning that the $\bm{A}$-field varies slowly enough with position that any changes in $\nabla\H{g}^R$ induced by the field can be neglected.
When the field is small, we can then perform a leading-order series expansion, and obtain
\begin{equation}
  \label{eq:series-expansion}
  \H{g} \approx \H{g}_0 + e\B{v}\cdot\B{A} \frac{\partial g_0}{\partial \epsilon}.
\end{equation}

The electric current can be calculated from the direction-dependence of the Keldysh propagator~$\H{g}^K$,
\begin{equation}
  \B{J} = -\frac{1}{8} N_Fe \int_{-\infty}^{+\infty} \d\epsilon\, \re \tr \langle \B{v} \H{\tau}_3 \H{g}^K \rangle,
\end{equation}
where $N_F$ is the normal-state density of states at the Fermi level, and the angle brackets denote an average over the Fermi surface.
Using the identities $\H{g}^K = \H{g}^R \H{h} - \H{h} \H{g}^A$ and $\H{g}^A = -\H{\tau}_3 \H{g}^{R\dagger} \H{\tau}_3$, and assuming that the distribution function has only an energy mode~$\H{h} = h(\epsilon) \H{\tau}_0$, the above can be rewritten as
\begin{equation}
  \B{J} = -\frac{1}{4} N_Fe \int_{-\infty}^{+\infty} \d\epsilon\, h(\epsilon) \re \tr \langle \B{v} \H{\tau}_3 \H{g}^R \rangle.
\end{equation}
If we have a system where no current flows in the absence of fields, then substituting in \cref{eq:series-expansion} yields to linear response
\begin{equation}
  \B{J} \approx -\frac{1}{4} N_Fe^2 \int_{-\infty}^{+\infty} \d\epsilon\, h(\epsilon) \, (\partial/\partial\epsilon) \re \tr \langle (\B{A}\cdot\B{v})\B{v} \H{\tau}_3 \H{g}^R \rangle.
\end{equation}
We now perform the averaging over the Fermi surface.
Noting that $\langle (\B{A}\cdot\B{v})\B{v} \rangle = (v_F^2/3)\B{A}$, while $(N_F/2) \re \tr \langle \H{\tau}_3 \H{g}^R \rangle = N(\epsilon)$ is just the density of states, the result is simply
\begin{equation}
  \B{J} = -\frac{1}{6} v_F^2e^2 \B{A} \int_{-\infty}^{+\infty} \d\epsilon\, h(\epsilon) \, (\partial N/\partial\epsilon).
\end{equation}
We may then perform an integration by parts,
\begin{equation}
  \B{J} = -\frac{1}{3} v_F^2e^2 \B{A} \left[ N_F - \int_{-\infty}^{+\infty} \d\epsilon\, \left(-\frac{\partial f}{\partial\epsilon}\right) \, N(\epsilon) \right],
\end{equation}
where we used $h(\pm\infty) = \pm 1$ and $N(\pm\infty) = N_F$ to evaluate the boundary terms, and $f(\epsilon) = [1-h(\epsilon)]/2$ to rewrite the distribution function as actual electronic occupation numbers.
This describes a local response $\B{J} = -K\B{A}$, where the kernel is
\begin{equation}
  K = \frac{1}{3} v_F^2e^2 \left[ N_F - \int_{-\infty}^{+\infty} \d\epsilon\, \left(-\frac{\partial f}{\partial\epsilon}\right) \, N(\epsilon) \right].
  \label{eq:kernel}
\end{equation}
This reproduces the standard textbook results~\cite{Schrieffer1999,Tinkham2004}, but has been derived entirely within the quasiclassical formalism.
Note that in reality, a clean system exhibits a nonlocal electromagnetic response; the local relation $\B{J} = -K\B{A}$ stems from the long-wavelength approximation made earlier.
If one instead rigorously calculates the full solution to the Eilenberger equation in a gauge field~$\B{A}$, one finds a very similar nonlocal relation
\begin{equation}
  \B{J}(z) = -K \langle \B{A} \rangle_z = -\frac{K}{d} \int\limits_0^d\d z'\,\B{A}(z'),
  \label{eq:nonlocal}
\end{equation}
where the clean-limit linear-response kernel~$K$ is the same as we derived within the Doppler approximation $\epsilon \rightarrow \epsilon + e\B{v}\cdot\B{A}$~\cite{Zaikin1982}.
We will not repeat the calculations shown in \eg Ref.~\cite{Zaikin1982}, but have now motivated why \cref{eq:nonlocal} together with \cref{eq:kernel} can describe the linear magnetic response of a clean system.

%-------------------------------------------------------------------------------%
%                                  BIBLIOGRAPHY                                 %
%-------------------------------------------------------------------------------%

% Bibtex file
\bibliography{references}

% Bibtex refs
% [include bbl-file here]